\documentclass[pra,10pt,tightenlines,amsmath,amssymb,amsfonts,twocolumn,superscriptaddress
,nofootinbib
]{revtex4}

\usepackage{natbib}
\usepackage{amsthm}
\usepackage{graphicx}
\usepackage{color}
\usepackage{bm}
\usepackage{mathptmx}

\graphicspath{{graphics/}}

\usepackage[colorlinks=false, linkcolor=blue, citecolor=magenta]{hyperref}

\newcommand\arrow\vec

\newtheorem*{definition}{Definition}

\DeclareMathOperator{\tr}{tr}

\def\system#1{\mathcal{#1}}
\def\const{\text{(const.)}}

\def\interact{\text{int}}
\def\lab{\text{lab}}

\def\avg#1{\left\langle {#1} \right\rangle}

\def\abs#1{\left\lvert{#1}\right\rvert}

\def\id{I}

\def\1{\mat{\id}}

\def\mat#1{\bm{\mathrm{#1}}}

\renewcommand{\vec}[1]{\bm{\mathrm{#1}}}

\def\ananke{\text{K}}
\def\chronos{\text{C}}

\begin{document}

\title{Clocks and Relationalism in the Thermal Time Hypothesis}

\author{Nicolas C. Menicucci}
\affiliation{Perimeter Institute for Theoretical Physics, Waterloo, Ontario N2L 2Y5, Canada}

\author{S. Jay Olson}
\affiliation{School of Mathematics and Physics, The University of Queensland, Saint Lucia, QLD 4072, Australia}

\author{Gerard J. Milburn}
\affiliation{Centre for Engineered Quantum Systems, The University of Queensland, Saint Lucia, QLD 4072, Australia}

\date{August 3, 2011}

\begin{abstract}
The Thermal Time Hypotheis (TTH) has been proposed as a general method for identifying a time variable from within background-free theories which do not come equipped with a pre-defined clock variable.  Here, we explore some implications of the TTH in an entirely relational context by constructing a protocol for the creation of ``thermal clocks'' from components of a large but finite quantum mechanical system.  The protocol applies locally, in the sense that we do not attempt to construct a single clock describing the evolution of the entire system, but instead we construct clocks which describe the evolution of each subsystem of interest.  We find that a consistency condition required for the evolution of our clocks is operationally equivalent to the general relativistic Tolman-Ehrenfest relation for thermal equilibrium in a static gravitational field but without the assumption of gravity or a metric field of any kind.
\end{abstract}

\maketitle

\section{Introduction}

Two of the most striking theoretical properties of general relativistic physics are the deep connection to thermodynamics and the unique role of time.  The connection to thermodynamics runs very deep indeed, starting historically with the realization that black holes obey laws in analogy to the laws of thermodynamics~\cite{Bekenstein1973}, followed by the theoretical discovery that black holes are \emph{literally} thermodynamic entities, described by quantities like entropy and temperature~\cite{Hawking1975}.  Ultimately, Einstein's equations themselves were derived as an equation of state~\cite{Jacobson1995}, and even Newtonian gravity has been derived as an entropic force~\cite{Verlinde2011}.

The role of time in relativistic physics is also fundamentally distinct from pre-relativistic physics~\cite{Kroes1984, Misner1973}.  The principle that all predictions should be free of reference to coordinates leads to a relational picture of physics, where a theory no longer predicts the evolution of variables with respect to a time variable $t$ through the functions $A(t)$, $B(t)$, $C(t)$, etc.,\ and instead a theory must predict how dynamical variables evolve with respect to one another, through functions such as $A(B)$, $C(B)$, etc.,\ where in this case $B$ would play the role of a ``clock variable''~\cite{Bergmann1961, Connes1994}. The issue is complicated by the fact that no single variable is generally accessible to play the role of clock for all observers in a general spacetime.  The puzzle then becomes how, in the general case, to identify the appropriate clock variable which ``plays the role of time'' for any given system, since the mechanical theory treats all variables equally, and it is known that different choices of clock variable can give rise to very different qualitative descriptions of the dynamics~\cite{Albrecht1995, Albrecht2008}.  This feature of GR is often hidden when working in a given background metric, where convenient coordinate systems may be chosen to correspond to the behavior of local measuring devices (i.e. gauge fixing), but it is nevertheless a fundamental property in a general setting, and coming to terms with its implications was one of the key insights in the development of GR~\cite{Norton1987}.

One idea which connects these two issues is known as the Thermal Time Hypothesis~(TTH)~\cite{Rovelli1993,Connes1994}.  The idea is that a fully relational theory should not be expected to identify its own time variable at the level of mechanics and that this should instead be the role of thermodynamics.  (Throughout this article, we use the quantum mechanical formalism to describe statistical states, but the TTH can just as well be applied to classical systems~\cite{Rovelli1993}. Our results do not depend on this distinction.)  The motivation for this idea comes from noticing that a fully thermalized state $\rho \propto {e^{- \beta H}}$ contains complete information on the dynamics of the system. That is, the Hamiltonian---and thus the time flow---of the system can be identified, up to an overall constant, by~$- \ln \rho$. The TTH takes this a step further and asserts that \emph{any arbitrary state} $\rho$ defines a time---the thermal time---given by the flow of $H_\rho = - \ln \rho$ on the observables, with respect to which the state~$\rho$ appears thermal.  A ``thermal clock,'' then, is a physical system prepared in a known initial state for which some observable (or some function of observables) increases with the flow of thermal time. It is with respect to these physical observables that the evolution of other systems may be compared~\cite{Connes1994}.

The full utility and role of ``thermal time'' is unknown at present. For example, it is not known how the time flow measured by an ordinary wristwatch is related to the thermal time of its owner---they cannot be identical time flows unless the owner is long dead (i.e.,~thermalized).
Nevertheless, the TTH has been explored in several contexts and shown to agree with known physics.  For example, in a classical radiation-filled cosmology, it selects out the expected Friedmann time~\cite{Rovelli1993a}, and in the quantum Minkowski vacuum restricted to the Rindler wedge it selects out a flow on the algebra of observables which corresponds to the trajectory of accelerated observers~\cite{Connes1994}. The TTH therefore recovers the Unruh-Davies effect~\cite{Davies1975,Unruh1976} in a manner complementary to the standard technique: the thermal state of the field in the Rindler wedge is taken as the starting point, with accelerated trajectories recovered as the corresponding flow of time.

In the above work, the objective has been to recover a single flow of time, but the introduction of a relational thermal clock has always been left as an implicit step in the framework.  Here, we explore the issue of thermal clocks explicitly by detailing a protocol for their construction, starting with a general statistical quantum state~$\rho$ describing a generic large but finite system~$\system S$.  In particular, we emphasize the relational character of the formalism, such that dynamical predictions come from the comparison of subsystems of $\system S$ to one another.

By partitioning $\system S$ into subsystems, we can introduce multiple thermal clocks, each describing the local dynamics of the system.  Within this construction, we find an inherent freedom to rescale the time flow locally. This can be understood by noting that the TTH effectively fixes~$H/T$ for each subsystem (with $H$ the Hamiltonian and $T$ the temperature), but it does not fix a scale for either $H$ or $T$ independently.  Thus, while the total system is by definition in equilibrium with respect to the thermal time flow, it is entirely consistent to have different temperatures for different subsystems.  We find that the consistency condition between these temperatures and the local time flow is operationally equivalent to the general relativistic Tolman-Ehrenfest~(TE) relation for temperature variation in a static gravitational field~\cite{Tolman1930a,Tolman1930}. That is, we obtain the relation~$T_A ds_A = T_B ds_B$, where $ds_A$ and $ds_B$ are local time intervals measured by our local, thermal clocks within two different subsystems, $\system A$ and $\system B$. We thus find that the TTH, when applied locally in a fully general, relational context and without appeal to the special properties of any specific system (in particular, without the assumption of any gravity or metric field), has a form of time dilation encoded within it and, in a sense, mimics the thermodynamic effects of a static spacetime metric.
 
Previous work has studied the TTH in connection with the TE relation~\cite{Rovelli2011} but from quite a different perspective than the one used here. Specifically, Ref.~\cite{Rovelli2011} starts with the assumption of a static gravitational field and uses the TTH and the equivalence principle to derive the TE relation, with the local temperature functioning as the ``speed of (thermal) time.'' In contrast, here we introduce no metric field at all and yet still recover a relation that is operationally equivalent to the TE relation and holds for arbitrary choices of initial subsystem temperature. This relation emerges naturally as a consistency condition that must be satisfied by local thermal clocks.

In Sec.~\ref{sec:objectiveH}, we begin by describing the construction of thermal clocks while retaining reference to a background time variable and an objective Hamiltonian governing the system.  This will ease the transition to the context of Sec.~\ref{sec:TTHandTE}, where we discuss the construction of thermal clocks in which the \emph{only} flow of time is that provided by the TTH itself. Therein we also discuss the physical interpretation of the temperature-scaling freedom and its relation to the perception of symmetry.  Sec.~\ref{sec:discussion} discusses the implications of our results.

\section{Objective Hamiltonian}

\label{sec:objectiveH}

While we wish eventually to apply the TTH to cases in which the dynamics are not independently specified, we must be sure that it succeeds in the cases in which the dynamics \emph{are} specified.  The first case we will consider, then, is that of a known Hamiltonian governing the dynamics of a finite-dimensional system. This will illustrate how the TTH picks out the correct Hamiltonian up to an overall constant (the temperature).  It has the further use of showing that the temperature-dependence goes away when one considers how such time flow would be \emph{measured physically}, which necessarily requires the construction of a clock.

\subsection{Dynamics from equilibrium}

We start with a nonrelativistic system~$\system S$ governed by a Hamiltonian~$H$ and having an associated time variable~$t$.  We assume that the system has a large but finite dimension.  We now imagine that the system is weakly coupled to a much larger system---a bath~$\system B$---and that the total system-bath combination~$\system S \otimes \system B$ is isolated and evolves according to some total Hamiltonian
\begin{align}
\label{eq:Htotal}
	H_{\text{total}} = H + H_{\text{bath}} + H_\interact\;.
\end{align}
``Weakly coupled'' means, in physical terms, that the interaction hamiltonian~$H_\interact$ can be ignored on timescales~$t \ll t_\lab := \| H_\interact \|^{-1}$ (with~$\hbar = 1$), where $\| \cdot \|$ is the operator norm (largest singular value) of its argument, which in this case corresponds to the largest interaction energy eigenvalue.  We call the timescale~$t_\lab$ the ``laboratory'' timescale because it is the timescale over which experiments can consider the evolution of~$\system S$ to be faithfully described solely by the Hamiltonian~$H$.  We want $t_\lab$~to be large in order to do meaningful experiments, which means that $\| H_\interact \|$~should be small (weak coupling).  In contrast, we want the system and bath to have interacted for a time~$t \gg t_\lab$ so that we can apply the results of Ref.~\cite{Popescu2006}, which allow us to claim that the state of~$\system S$ will be very close to a thermal state when we ignore (i.e.,\ trace out)~$\system B$.

From this point on, we ignore the bath.  The state of~$\system S$ will be very close to~\cite{Popescu2006}
\begin{align}
\label{eq:rhothermal}
	\rho = \frac {e^{-\beta H}} {\tr e^{-\beta H}}\;,
\end{align}
where $\beta$~is the inverse temperature and is a function of the total average energy in the system.  Applying the TTH to this case is rather trivial.  The \emph{thermal Hamiltonian} (i.e.,\ the dynamics that are picked out by the TTH) associated with the state~$\rho$ is given by
\begin{align}
\label{eq:TTHonrhothermal}
	\beta' H_\rho := -\ln \rho = \beta H + \const\;,
\end{align}
where the constant represents an overall energy shift that can be ignored, and the arbitrary inverse temperature~$\beta'$ is a free parameter that is not fixed by the TTH.  (The TTH technically, therefore, picks out an equivalence class of thermal Hamiltonians.)  Without loss of generality, then, we can say that
\begin{align}
\label{eq:HthermalproptoH}
	H_\rho = \alpha H\;,
\end{align}
where
\begin{align}
\label{eq:alphadef}
	\alpha = \frac {\beta} {\beta'}\,.
\end{align}
Thus, the thermal Hamiltonian identifies dynamics that are correct up to an overall constant~$\alpha$ equal to the ratio of the actual inverse temperature~$\beta$ to the freely assigned inverse temperature~$\beta'$.

\subsection{Temperature dependence}

It might seem problematic that there is an arbitrary overall scaling of the thermal Hamiltonian that depends on an arbitrarily chosen temperature, but let's look at this issue a little more closely.  Specifically, let's be clear about the physical situation and what is happening.  An agent stumbles upon a system that is relatively isolated on the timescales he has to work with but which has been interacting very weakly with a huge reservoir for a long time.  The agent therefore, with good reason~\cite{Popescu2006}, believes that the system is in thermal equilibrium with the bath and is itself in a thermal state.  She does not know what this state is, however---all she currently has is good reason to believe that the system is in an equilibrium state.  Now imagine that she acquires knowledge of what the state is.  This could be through her possessing a useful theory of the physical world, by being told by another agent what the state is, or possibly by doing tomography on a large set of identically prepared systems (which requires that such a set is available and that her measurements are sufficiently precise).  She now has a state~$\rho$ that she believes to be a thermal equilibrium state.  She then postulates that the temperature of this state is~$\beta'$ and uses the TTH to calculate a thermal Hamiltonian~$H_\rho$.  Since in general $\beta' \neq \beta$, she cannot recover~$H$ directly.  So how can she check whether $H_\rho$ describes the dynamics correctly?

One common---albeit unfounded---objection to the TTH based on the fact that thermal states don't evolve~\cite{Connes1994}.  While a true statement, this is irrelevant as long as the system can be reprepared in a new state.  Such preparations might include small perturbations around thermality, as discussed in Ref.~\cite{Connes1994}, but this is not the only possibility; an entirely new state might be prepared instead.  Either way, the upshot is the same: the TTH uses the notion of equilibrium, along with a state that is believed to satisfy the equilibrium conditions, to define dynamics for the system; these dynamics are tested by making observations on a newly prepared state that is \emph{not} an equilibrium state.

Therefore, armed with a system believed to be in equilibrium and a state~$\rho$ believed to describe it, our agent reprepares the system in some new state~$\sigma$ that does not commute with~$\rho$ and then makes measurements of some ovservable~$A$ at various times, comparing the results to the predictions given by the use of~$H_\rho$ as the system's Hamiltonian.  To predict expectation values of~$A$, she would use the Heisenberg equation of motion with the thermal Hamiltonian, predicting that
\begin{align}
\label{eq:thermalHeis}
	\frac {d} {ds} A_\rho (s) &= i [H_\rho, A_\rho(s)]\;,
\end{align}
with $A_\rho(0) = A$.  The subscript~$\rho$ indicates that the state~$A_\rho(s)$ is the Heisenberg-picture observable corresponding to~$A$ that is predicted using~$H_\rho$, and $s$~is called the \emph{thermal time variable} to distinguish it from the physical time variable~$t$. The solution to Eq.~\eqref{eq:thermalHeis} is
\begin{align}
\label{eq:Arho}
	A_\rho(s) = U_\rho^\dag(s) A U_\rho(s)\;,
\end{align}
where
	$U_\rho(s) = e^{-iH_\rho s}$
is the (thermal) time-evolution operator.  Clearly, testing this evolution requires tracking the evolution of~$A$ \emph{with respect to~$s$}.  But this raises the important question, \emph{How can our agent measure~$s$?}  Clearly, she needs a \emph{clock} of some sort.  We will show below how she can build such a clock, but for the moment we will discuss what she should expect to see once she has such an $s$-measuring clock.

The true Heisenberg equation of motion for the observable is
\begin{align}
\label{eq:trueHeis}
	\frac {d} {dt} A (t) &= i [H, A (t)]\;,
\end{align}
where the lack of subscript on $A(t)$ means that this Heisenberg-picture observable is defined using the actual time evolution given by~$H$, and again we have that~$A(0) = A$.  The solution is
\begin{align}
\label{eq:Aoft}
	A(t) = U^\dag(t) A U(t)\;,
\end{align}
where
	$U(t) = e^{-iHt}$
is the physical time-evolution operator.  The connection between the two evolutions is made by noticing that
\begin{align}
\label{eq:compareUs}
	U_\rho(s) = e^{-iH_\rho s} = e^{-iH \alpha s} = U(\alpha s)\;,
\end{align}
where $\alpha$~is defined in Eq.~\eqref{eq:alphadef}, thus revealing that the thermal time-evolution operator acts just like the physical one, except that the time variables are rescaled by a temperature-dependent constant:~\mbox{$t = \alpha s$}.  Thus,
\begin{align}
\label{eq:compareAs}
	A_\rho(s) = A(\alpha s)\;.
\end{align}
If a different temperature~$\beta'' \neq \beta'$ had been assumed instead, then this change could have been completely accounted for in the analysis by rescaling the thermal time variable.  \emph{Changing the arbitrary temperature used in applying the TTH corresponds exactly to rescaling the thermal time variable.}  This is the only effect.  We now show that the effect is not detectable under normal circumstances.

\subsection{Clocks}
\label{subsec:clocks}

We mentioned in the last subsection that should an agent wish to test the time evolution she calculates using the TTH, she will need to track measurements with respect to the thermal time~$s$.  But the ``thermal time'' is simply a \emph{coordinate} and as such is not by itself physical.  The same is true with the ``physical time''~$t$, which itself is also just a coordinate (despite the nomenclature).  To actually measure changes with respect to either time variable, one needs to construct a clock.

\begin{definition}
A \emph{clock}, for our purposes, is a system that is prepared in a known state and allowed to evolve with known dynamics during which the expectation value of some observable of that system increases monotonically with the time variable (and this scaling function is known).
\end{definition}

No clock will function forever, and no clock can ever be perfect even for a limited time~\cite{Milburn1991,Milburn2005}.  Still, one can construct a clock that is useful up to a given accuracy and for some finite duration. (We will use a harmonic oscillator over half a period as an example below.) It is important that the clock function as a closed system (i.e.,~with unitary evolution) up to the desired level of accuracy. In particular, it must not interact with the system whose evolution is to be tracked. While it is true that some clocks are easier to use and may function with more accuracy and for longer times than others, we will make no mention of the term ``good clock'' as it is sometimes used in the literature~\cite{Albrecht2008} since anything else (i.e.,\ a ``bad clock'') for our purposes is not a clock at all.

Imagine now that the agent observing~$\system S$ determines that the equilibrium state~$\rho$ can be written as
\begin{align}
\label{eq:clocksys}
	\rho = \rho_C \otimes \rho_O\;,
\end{align}
meaning that the agent observes that the system behaves as if it were composed of two subsystems~$\system C$ and~$\system O$.  Most states cannot be written in this form, so if $\rho$~can be so decomposed, the tensor-product decomposition in Eq.~\eqref{eq:clocksys} is not arbitrary but is, rather, fixed by the state itself. Applying the TTH to this state gives
\begin{align}
\label{eq:TTHonclock}
	H_\rho = H_{\rho_C} \otimes \id + \id \otimes H_{\rho_O}\;,
\end{align}
where~$\beta' H_{\rho_C} = -\ln \rho_C$, $\beta' H_{\rho_O} = -\ln \rho_O$.

If the state~$\rho$ truly is an equilibrium state for the dynamics~$H$, then ignoring total shifts in energy, we can write
\begin{align}
\label{eq:clocksysfromH}
	\rho &= e^{-\beta H} = e^{-\beta H_C} \otimes e^{-\beta H_O}\;,
\end{align}
and therefore~$H_{\rho_C} = \alpha H_C$, and~$H_{\rho_O} = \alpha H_O$.  The crucial thing to notice is that \emph{the temperatures of the subsystems are the same}, a requirement we will call \emph{Isothermality}.  Isothermality is often understood as a necessary consequence of the Zeroth Law of Thermodynamics, which states that two systems in equilibrium with a third are in equilibrium with each other. (In light of relativity, however, the Zeroth Law need not imply Isothermality~\cite{Tolman1930}, a fact we will ignore for now but which will be crucial later.) Thus, even though there is no direct interaction (Hamiltonian term) linking~$\system C$ and~$\system O$, they have equilibrated with each other by virtue of their mutual contact with the bath~$\system B$.  This means their temperatures should be equal.

The agent can use this fact to construct a clock from~$\system C$, which she can use to test the dynamics she predicts for the object~$\system O$.  Applying the TTH to Eq.~\eqref{eq:clocksysfromH}, along with Eq.~\eqref{eq:compareUs}, gives
\begin{align}
\label{eq:Uclocksys}
	U_\rho(s) &= U_{\rho_C}(s) \otimes U_{\rho_O}(s) \nonumber \\
	&= e^{-i H_{\rho_C} s} \otimes e^{-i H_{\rho_O} s} \nonumber \\
	&= e^{-i H_C \alpha s} \otimes e^{-i H_O \alpha s} \nonumber \\
	&= U_C(\alpha s) \otimes U_O(\alpha s)\;,
\end{align}
where~$U_{\rho_C}$ and~$U_{\rho_O}$ are defined by Eq.~\eqref{eq:compareUs} for the individual systems~$\system C$ and~$\system O$, respectively, and $\alpha$~is defined in Eq.~\eqref{eq:alphadef}.  As a concrete example for the clock, consider the case in which a large number of the lowest energy eigenvalues of~$H_C$ are nondegenerate and evenly spaced.  Up to some energy scale (which we assume is large enough for our purposes), this system behaves like a harmonic oscillator with some frequency~$\omega$.  Such a system can be used as a clock for measuring durations $\Delta t \ll \pi \omega^{-1}$ (half of one period).  All that is required is to prepare the oscillator in a coherent state and let it go.  Coherent states behave like noisy states of a classical oscillator, and thus
\begin{align}
\label{eq:avgq}
	\avg {q(t)} = q_0 \cos (\omega t + \varphi)\;,
\end{align}
where $q_0$~is an overall amplitude, and $\varphi$~is the phase parameter, which depends on the initial average position and momentum.  For large-amplitude coherent states (those with an initial displacement whose magnitude is much larger than ground-state uncertainty), and with~$\varphi = -\pi/2$, the measured value of~$q(t)$ (up to some small uncertainty) increases monotonically with~$t$ for~$\abs t < \pi/2 \omega$ (i.e.,\ for half of one period) and can be used to make an observable that approximately measures~$t$:
\begin{align}
\label{eq:tfromq}
	\tau(t) = \frac 1 \omega \left[ \cos^{-1} \left(\frac {q(t)} {q_0}\right) - \varphi \right]\;.
\end{align}
The observable~$\tau(t)$ tracks the physical time~$t$ (i.e.,\ $\avg {\tau(t)} = t$ for values of~$t$ limited to a single half-period) by virtue of the system's known dynamics and initial state.  This is not the only type of system that can be used as a clock.  (Other possibilities include, for example, a large magnetic dipole precessing in a magnetic field~\cite{Poulin2006}.)  We use the harmonic oscillator as a concrete example. We only consider unitary clocks for the moment because calibrating a nonunitary clock (e.g.,~atomic decay clock) would require knowledge of the coupling between the clock and its environment, and we do not have this luxury within the limited-knowledge paradigm of the TTH.

An agent using the TTH will predict
\begin{align}
\label{eq:thermalclocktime}
	\tau_\rho(s) &= \tau(\alpha s)\,, \\
\label{eq:thermalobservable}
	A_\rho(s) &= A(\alpha s)\,,
\end{align}
i.e.,\ the clock and object will evolve like they should but at a different overall rate in thermal time~$s$ than they do in physical time~$t$.  The question, however, is whether the evolution \emph{with respect to the clock reading} as predicted by the TTH is the same as that predicted by using the real dynamics instead.  It is intuitive that this is the case, but we shall prove it rigorously.

At physical time~$t$, the clock reads~$\bar \tau$, and the observable for~$\system O$ is~$A[\tilde t(\bar \tau)]$, where we use $\tilde t(\bar \tau)$~to denote the function used to estimate the physical time~$t$ from a clock reading of~$\bar \tau$.  This function satisfies~$\tilde t[\tau(t)] = t$ up to an uncertainty assumed small enough to neglect.  The agent, however, will declare that the clock reading of~$\bar \tau$ corresponds to a thermal time of $\tilde s(\bar \tau)$, which must satisfy~$\tilde s[\tau_\rho(s)] = s$.  Using these facts, along with Eq.~\eqref{eq:thermalclocktime}, gives
\begin{align}
	 \tilde t[\tau(\alpha s)] = \alpha s = \alpha \tilde s[\tau_\rho(s)] = \alpha \tilde s[\tau(\alpha s)]
\end{align}
for all thermal times~$s$ in the usable range of the clock.  We can therefore identify the functions
\begin{align}
\label{eq:clockestimates}
	\tilde t(\bar \tau) = \alpha \tilde s(\bar \tau)\,.
\end{align}
At a clock reading of~$\bar \tau$, the agent will infer a thermal time of~$\tilde s(\bar \tau)$ and predict that the observable will have evolved to~$A_\rho[\tilde s(\bar \tau)]$. Using Eqs.~\eqref{eq:thermalobservable} and~\eqref{eq:clockestimates}, this becomes
\begin{align}
	A_\rho[\tilde s(\bar \tau)] = A[\alpha \tilde s(\bar \tau)] = A[\tilde t(\bar \tau)]\;,
\end{align}
which, as evidenced by the right-hand side, is exactly what is predicted by using the correct dynamics and physical time.  The dependence on~$\alpha$---and hence on the arbitrary temperature~$\beta'$---has been completely eliminated.

\subsection{Note on ``objectivity''}
\label{subsec:objectivity}

Throughout this section we have used phrases like ``objective Hamiltonian'' and ``real dynamics'' and ``physical time.'' The ontological status bestowed by this kind of terminology has been useful so far because it provides an authoritative goal for the TTH to strive to achieve when applied to well-understood systems---i.e.,\ we have known dynamics that the TTH should be able to reproduce. We want to theoretically examine the TTH in this context as a first step, and this is why we have employed such authoritative terminology. But our ultimate goal is to apply the TTH to systems in which there is no objective flow of time---Rovelli's original purpose~\cite{Connes1994}---or to those in which there is an ambiguity as to which agent's assessment of time flow is correct. For these purposes, such terminology is misleading, so we would like to move away from it.

At this point we would like to point out that everything that was claimed about an ``objective'' Hamiltonian, dynamics, time, etc.\ can be equally well understood in terms of quantities whose only claim to fame is that they have been \emph{verified} by some particular observer. If all observers agree, then there is no harm in calling such findings ``objective,'' but if other observers can make other assessments of the same physical situation and be equally correct in terms of their observations, then we have an interesting situation. In fact, this is exactly the situation in relativity: observers in different states of motion and/or different gravitational potentials make different assessments of the same physical situation. Yet they all believe themselves to be correct, and they are all justified in believing so by observation. The theory of relativity is nothing more than the resolution to the paradox of how they can all be right even though they disagree.

From now on it should be understood that whenever we speak of the ``correct'' dynamics, temperature, time, etc.\ what we mean is such quantities that have been verified (or could in principle be verified) to be correct by some particular agent. In fact, predictions for these quantities could have been the result of a prior application of the TTH. Our results do not depend on whether these are the real state of affairs or just what some agent believes, as long as they are \emph{verifiable} by that agent. (The reader may wish to review this section in light of this relaxed understanding of ``objective,'' as well.)

\section{Temperature-Assignment Freedom and the Tolman-Ehrenfest Relation}
\label{sec:TTHandTE}

\subsection{Violation of Isothermality in general relativity}

The temperature of an extended system in thermal equilibrium in a static gravitational field varies with the local gravitational potential~\cite{Tolman1930a,Tolman1930}. An intuitive way to understand this is to consider that the energy of a particle increases as it falls and decreases as it rises. Thus, for the energy exchange to be balanced as required by the conditions of equilibrium, the upper portion must be slightly cooler than the lower portion. Otherwise there would be a net energy flow downward through local thermodynamic exchange processes. Violation of Isothermality is possible---and generally expected---for an extended system in a gravitational field.

The Tolman-Ehrenfest~(TE) relation describes the relationship between temperatures of two regions of a composite systems that has come to equilibrium in the presence of a static gravitational field~\cite{Tolman1930}. The result is that:
\begin{align}
\label{eq:Tolmanrel}
	T(x) \sqrt{g_{00}(x)} = \const\;,
\end{align}
where the metric is
\begin{align}
\label{eq:Tolmanmetric}
	ds^2 = g_{00} dt^2 + g_{jk} dx^j dx^k\;,
\end{align}
with all metric components independent of~$t$.  While this relation is written in terms of temperatures and metric components, its operational content can be described in its entirety as a relation between the local temperature~$T(x)$ and the  tick rate of comoving local clocks. This is easily seen by considering an observer on a worldline of constant~$x$. A single tick of his clock~($ds$) is given in terms of the ticks of the global coordinate time~($dt$) by the simple formula $ds = \sqrt{g_{00}} dt$.  A physically meaningful comparison, then, is between the rate of proper time flow for observers at two different locations.  Agents~$A$ and~$B$ located at two different fixed~$x$ coordinates have watches that measure proper times~$s_A$ and~$s_B$, respectively.  Equation~\eqref{eq:Tolmanrel} can therefore be expressed as
\begin{align}
\label{eq:Tolmands}
	T_A ds_A = T_B ds_B\;,
\end{align}
whose operational content can be described without appeal to a metric at all.
\begin{definition}
\emph{Operational content of the TE relation:} the times measured by each observer's wristwatch flow at rates inversely proportional to respective local temperature measurements, assuming the composite system being measured is in thermal equilibrium.
\end{definition}
\noindent Notice that this statement makes no reference to general relativity. It only involves temperatures and clock rates, along with the notion of thermal equilibrium. On the other hand, the \emph{applicability} of the TE relation, as it was originally derived, is limited to equilibrium in a static gravitational field with observers at fixed spatial locations. This is where general relativity comes into play in its original derivation. But the operational content of the TE relation, as identified in Eq.~\eqref{eq:Tolmands} and explained in the Definition above, can be stated and considered independently of this assumption.

In what follows, we will show that the operational content of the TE relation shows up in other contexts besides just static gravitational fields. In fact, we will derive Eq.~\eqref{eq:Tolmands} directly from three components: (1)~the TTH, (2)~relationalism, and (3)~the assignment of different local temperatures to subsystems within a fully thermalized, extended system. In doing so, we will make no reference to general relativity. It is this crucial conceptual difference---separating the operational content of the TE relation from its gravitational roots---that distinguishes this work from Ref.~\cite{Rovelli2011} and also from other information-theoretic derivations of the TE relation~\cite{Daffertshofer2007}.

\subsection{Violation of Isothermality in the TTH}
\label{subsec:tempdiffTTH}

We wish to uplift the empirical content of the TE relation from its relativistic roots. To this end, we will discard everything we know about relativity, including the TE relation itself, and consider a toy example to which we will apply the TTH. We assume that there are two agents, Ananke and Chronos.\footnote{In primordial Greek mythology, Ananke is the goddess of fate, and Chronos is the god of time.} Both agents know that a given system is in thermal equilibrium and obtain its state~$\rho$ either through tomography, by theoretical calculation, or by being told enough reliable information about its statistics. Our agents both observe that the density matrix decomposes as
\begin{align}
\label{eq:rhotwosystems}
	\rho = \rho_A \otimes \rho_B\,,
\end{align}
which defines two independent subsystems~$\system A$ and~$\system B$, respectively. We further assume that
\begin{align}
\label{eq:rhoA12}
	\rho_A &= \rho_{A1} \otimes \rho_{A2}\,, \\
\label{eq:rhoB12}
	\rho_B &= \rho_{B1} \otimes \rho_{B2}\,,
\end{align}
where the sizes of the four Hilbert spaces need not be related to each other. This splitting will be useful for defining local clocks to test the dynamics predicted for subsystems~$\system A$ and~$\system B$. (Note that both Ananke and Chronos have access to all subsystems of the original composite system; these tensor-product decompositions have nothing to do with the agents.)

We used a similar starting point in Section~\ref{subsec:clocks}, from which we showed that one subsystem can serve as a local clock for the other and that the temperature assigned in applying the TTH is irrelevant operationally. This required assuming Isothermality. In this example, we will let one of our agents, Ananke, do this, while the other agent, Chronos, does not. Taking the discussion in Section~\ref{subsec:objectivity} seriously, we will not posit a separate ``objective'' dynamics and will instead assume that the assumptions given---i.e.,~overall thermal equilibrium and a given state---are correct. This is enough to ensure that the TTH is applicable. Our procedure instead will be to assume that each agent, in turn, is ``correct'' in their predictions and subsequently show how the other agent can be as well. If we can do this, then we will have shown that neither one has a monopoly on truth---i.e.,~either viewpoint is equally valid---and we will have necessarily developed a theory about how that can be the case.

For Ananke, there is only one temperature in question, whose inverse she calls~$\beta$. She chooses this temperature arbitrarily and uses it when applying the TTH to~$\rho$, thereby obtaining the Hamiltonian~$H^\ananke$ for the entire system (we use `$\ananke$' to label Ananke's quantities because `$A$' is already in use as a system label). This Hamiltonian satisfies
\begin{align}
\label{eq:Hrhoananke}
	\beta H^\ananke = -\ln \rho\,.
\end{align}
Because she assumes Isothermality, she also has the following relations:
\begin{align}
\label{eq:HABananke}
	\beta H^\ananke_A &= -\ln \rho_A\,,	\nonumber \\
	\beta H^\ananke_B &= -\ln \rho_B\,,
\end{align}
resulting in
\begin{align}
\label{eq:Hrelationananke}
	H^\ananke = H^\ananke_A \otimes \id + \id \otimes H^\ananke_B\,.
\end{align}
She can similarly define individual Hamiltonians for systems~$\system A_1$, $\system A_2$, $\system B_1$, and $\system B_2$. To test her dynamics, she reprepares the joint system in a new state~$\sigma$ that does not commute with~$H^J$. She then predicts the evolution of the expectation value of an observable~$O$ with the Heisenberg equation of motion, using~$t$ as the label for the only thermal time variable she has:
\begin{align}
\label{eq:Heisananke}
	\frac {d} {dt} O^\ananke (t) &= i [H^\ananke, O^\ananke (t)]\,.
\end{align}
She can test these dynamics by choosing any of the four subsystems to act a clock, as described in Section~\ref{subsec:clocks}, for any other the other three (or some combination thereof). The effect of assuming Isothermality when applying the TTH is that \emph{the entire overall system evolves synchronously with respect to thermal time~$t$}. Since, by assumption, Ananke has the correct state and is correct about it being in global thermal equilibrium, she must be able to make accurate empirical predictions using~$H^\ananke$ even if she has chosen~$\beta$ arbitrarily. (This is the main result of Section~\ref{subsec:clocks}.) Notice that this statement does not rely on Isothermality being correct because it does not include any reference to subsystems. Later on, we will have more to say about how this can be, but for the moment, let's see what Chronos obtains by forgoing Isothermality in his predictions.

Having chosen not to apply Isothermality---a decision that will be given possible motivation later---Chronos now has two independent temperatures to choose, whose inverses he calls~$\beta_A$ and~$\beta_B$. Having made this decision, Chronos is forgoing the ability for~$\system A$ (or any part thereof) to act as a clock for~$\system B$ and vice versa. This is because, as shown in Section~\ref{subsec:clocks}, the use of a TTH-constructed clock to track the TTH-predicted dynamics of a system requires that the temperatures used for applying the TTH to each of those systems initially were the same. He could get around this problem by making use of the ratio of~$\beta_A$ and~$\beta_B$ to scale the clock reading appropriately, but that doesn't change the basic fact that the TTH-derived clock and system are not necessarily expected to run synchronously unless Isothermality has been assumed.

This may seem like an unnecessary and unwelcome complication since one could argue that the ``real'' time is the global time derived by Ananke, which she calls~$t$. After all, our thermodynamic intuition for systems in thermal equilibrium is almost exclusively for those that also happen to be at a uniform temperature. But we cannot dismiss Chronos's choice so easily. In fact, we have to be true to our goal for the TTH, which is to have a flow of time \emph{emerge} from thermodynamic principles. As such, with temperature being a free parameter in the TTH, we can choose to be like Ananke and let emerge a single uniform time, the birthplace of which is assuming Isothermality; or we can choose to be like Chronos and apply the TTH separately to the two subsystems, therein choosing two different temperatures and obtaining \emph{local time flows} that are independent of each other (even if they are connected by a temperature-dependent scaling factor).

Chronos starts with the same~$\rho$ from Eq.~\eqref{eq:rhotwosystems} but does not apply the TTH to~$\rho$ directly. Therefore, he does not assign a global temperature for the entire system. Instead, he treats subsystems~$\system A$ and~$\system B$ independently and applies the TTH separately to~$\rho_A$ and~$\rho_B$, with his temperature choices noted above. As such, Chronos obtains the following thermal Hamiltonians:
\begin{align}
\label{eq:HABchronos}
	\beta_A H^\chronos_A &= -\ln \rho_A\,,	\nonumber \\
	\beta_B H^\chronos_B &= -\ln \rho_B\,.
\end{align}
Like Ananke, Chronos can verify the local evolution of subsystems~$\system A$ and~$\system B$ by splitting them each, separately, into clock and object, as in Section~\ref{subsec:clocks}. While he does not have the same freedom that Ananke did in choosing, say, a clock from~$\system A$ to track the evolution of an object in~$\system B$, he can still verify each subsystem's evolution \emph{locally}, and by the results of Section~\ref{subsec:clocks}, the predictions will check out even though his temperature choices are different from Ananke's. Notice that, despite forgoing the application of Isothermality globally, Chronos has nevertheless applied it locally to subsystems~$\system A$ and~$\system B$. This is required in order to ensure that his predicted evolution for those systems can be verified and makes sense because we want a notion of locality for the individual subsystems~$\system A$ and~$\system B$.

Comparing Eqs.~\eqref{eq:HABananke} with Eqs.~\eqref{eq:HABchronos}, Chronos's local Hamiltonians can be seen to relate to Ananke's by a local scaling factor:
\begin{align}
\label{eq:HABconnection}
	H^\chronos_A &= \alpha_A H^\ananke_A\,,	\nonumber \\
	H^\chronos_B &= \alpha_B H^\ananke_B\,,
\end{align}
where
\begin{align}
\label{eq:alphaAB}
	\alpha_A = \frac {\beta} {\beta_A} \qquad \text{and} \qquad
	\alpha_B = \frac {\beta} {\beta_B}\,.
\end{align}
Chronos might be tempted to try to write down a global Hamiltonian that includes both systems like Ananke did, such as $H^\chronos = H^\chronos_A \otimes \id + \id \otimes H^\chronos_B$, but this would not be proportional to~$H^\ananke$ because $\alpha_A \neq \alpha_B$. This would also be conceptually problematic because Chronos has only applied the TTH \emph{locally} to each subsystem, thereby obtaining two \emph{local} thermal time flows. He has not applied the TTH globally to~$\rho$ itself and thus, for Chronos, a notion of global time flow has not emerged. A global time variable is required in order to define a notion of simultaneity across the subsystems. Ananke can make this definition because she assumed Isothermality when applying the TTH globally, but Chronos cannot.

Just as ancient Greek mythological deities are now understood to be merely different aspects of the natural world, so do Ananke and Chronos in this discussion correspond to different approaches to TTH-based time evolution for the same system and equilibrium state. Both Ananke's and Chronos's thermal time coordinates are just that---coordinates, which themselves do not have physical meaning. As shown in Sec.~\ref{subsec:clocks} and reiterated many times above, the coordinates themselves must be eliminated from empirical predictions before such predictions can be considered physical. Since Ananke and Chronos each correspond to a different choice of temperature assignment for the TTH, and since a priori no particular assignment is blessed with being the ``correct'' assignment (see Sec.~\ref{subsec:objectivity}), we can use the coordinates defined by both agents within the same calculation.\footnote{If this seems objectionable, just consider calculations involving a static metric in relativity. These can involve a global coordinate time, which is often called~$t$, as well as a multitude of proper times corresponding to many observers, and these all can differ despite referring to the same physical objects.} Doing so will allow us to derive the consistency conditions linking the two viewpoints.

Chronos can therefore use Anake's global thermal time~$t$ as a bridge to compare his local times for the two systems, $s_A$ and~$s_B$. This works because Ananke and Chronos have used two different temperature assignments to make predictions about the dynamics of the same physical system. By the methods of Section~\ref{subsec:clocks}, Chronos's local thermal times relate to the Ananke's global time~$t$ through the respective relations
\begin{align}
	t = \alpha_A s_A\;, \nonumber \\
	t = \alpha_B s_B\;.
\end{align}
We can relax these conditions to allow for arbitrary shifting of the zeros of all three scales and still write
\begin{align}
	dt = \alpha_A ds_A\;, \nonumber \\
	dt = \alpha_B ds_B\;,
\end{align}
which, after eliminating~$dt$ and~$\beta$, give
\begin{align}
\label{eq:TTHtoTolman}
	T_A ds_A = T_B ds_B\;.
\end{align}
Notice that this has \emph{exactly the same physical content} as Eq.~\eqref{eq:Tolmands}, which was obtained in Ref.~\cite{Tolman1930} only by employing the full machinery of general relativity: namely that the time variable measured by a local clock (proper time) passes at a rate inversely proportional to the perceived temperature at that location when the overall state is in thermal equilibrium. The key result here is that using the principle of relationalism---which means that one must construct a clock to measure time---and the temperature-assignment freedom granted by TTH, we get a result that is physically equivalent to the TE relation.  Another way of looking at this is that given a theory that predicts an equilibrium state with different temperatures for different subsystems, applying the TTH with theory-assigned temperatures gives TE-style time dilation automatically. One important outstanding question is, \emph{Aside from gravity, why would anyone assign different temperatures to a composite system in equilibrium?} It is to this question---and one possible answer to it---that we now turn.

\subsection{A tale of two students: \\ temperature variation within equilibrium}
\label{subsec:exampletempvar}

We wish to make a connection from the abstract description in Sec.~\ref{subsec:tempdiffTTH} to concrete laboratory physics. To this end, here we describe a fictitious scenario starring Ananke and Chronos as experimental physics Ph.D.\ students.

Ananke's and Chronos's disagreements about how to assign temperatures to composite systems in equilibrium---specifically, whether Isothermality must be invoked or not---dates back to their days as Ph.D.\ students. Both studied under the supervision of Prof.\ Chaos, who was both cunning and unpredictable. Prof.\ Chaos was a quantum electrodynamics experimentalist, specializing in the precise engineering, control, and testing of quantum dots. One morning, Ananke and Chronos show up to the lab and discover a note left for them by the Professor, along with a lab notebook belonging to their fellow student, Gaia:

\begin{quotation}
\noindent Dear Ananke and Chronos,

As you both know, Gaia has been working on an experiment to characterize two quantum dots. She has already taken a substantial amount of data on the dots and recorded it in this notebook. Unfortunately, just after completing the data-taking session late last night, she fell ill and had to be rushed to the hospital. She is fine now, but in her hasty departure, the lab was left unlocked, and the experiment was vandalized. She is supposed to present a poster on it next week, but she did not have time to analyze the dynamics of the dots before they were damaged. All she has are data on the state of the dots.

She will be out for the rest of this week, and I will be away, as well, to work on Project Universe, but I am hoping that the two of you can find a way to use the data that Gaia collected to complete some sort of poster for the conference. If you are willing to help, both of your names will be added to the author list, and Gaia will be forever grateful.\\
\hspace*{10em} Chaos
\end{quotation}

\begin{figure}[t!]
\centering
\includegraphics[width=\columnwidth]{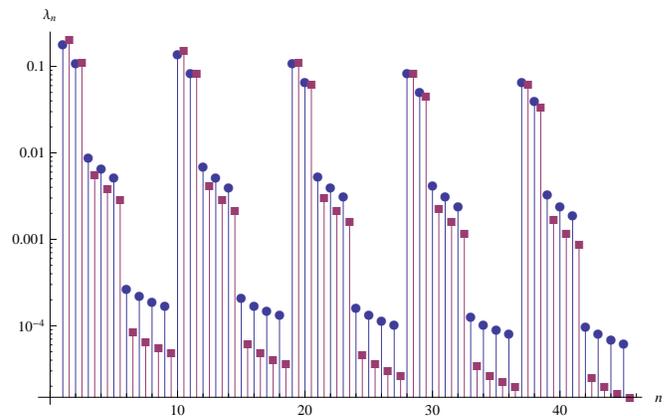}
\caption{\label{fig:dothistogram}Eigenvalues of the density matrix~$\rho$ for the two dots on a logarithmic scale. Those for dot~$\system A$ are indicated by the blue circles; those for dot~$\system B$, by the purple squares. The grouping of each spectrum into a five-fold repetition indicates that both~$\rho_A$ and~$\rho_B$ have a tensor-product decomposition. Visual similarity of the dots' spectra can be explained either as two different dots at the same temperature (Ananke) or as two identical dots at two different temperatures (Chronos). See text and Figs.~\ref{fig:levelsconstT} and~\ref{fig:levelsdiffT}.}
\end{figure}

With no access to the physical dots themselves, the two students have only the data in Gaia's lab notebook to work with. They open it up and discover that Gaia has taken an enormous amount of tomographic data on the two dots. Ananke gets to work writing a state estimation algorithm, while Chronos looks through the notebook for additional clues. Unfortunately, there is little recorded beyond just results of projective measurements made on each dot. The only thing Gaia notes is that the dots were at ``room temperature'' when the measurements were made. Chronos surmises that the dots were therefore in thermal equilibrium, with the bath being the laboratory itself. He notes that room temperature is around 300~K, but without a precise temperature given, they will have to use their best guess. Ananke finishes the algorithm, and the pair begin to input the data. After several hours, they get a result---a density matrix~$\rho_A$ for the dot Gaia has called~$\system A$ and another density matrix~$\rho_B$ for the dot Gaia labels as~$\system B$.

But what good are these states if what they are looking for is the dynamics of the dots (i.e.,~their respective Hamiltonians)? The students search the web for ``dynamics from equilibrium'' and discover Rovelli's writings on the TTH~\cite{Rovelli1993,Connes1994}. Excited by this idea, they diagonalize the density matrix for each dot and discover a very interesting pattern in their eigenvalues, which are plotted in Figure~\ref{fig:dothistogram}. Analysis reveals two interesting things about the spectra. First, the spectra of the two dots are visually similar in structure. They check this numerically, and it turns out that they are in fact proportional to each other on a logarithmic scale (up to normalization). Second, it turns out that the two spectra each reduce to a Cartesian product of two subspectra (on a linear scale), indicating that~$\rho_A = \rho_{A1} \otimes \rho_{A2}$ and~$\rho_B = \rho_{B1} \otimes \rho_{B2}$.

\begin{figure}[t!]
\centering
\fbox{\includegraphics[width=0.95 \columnwidth]{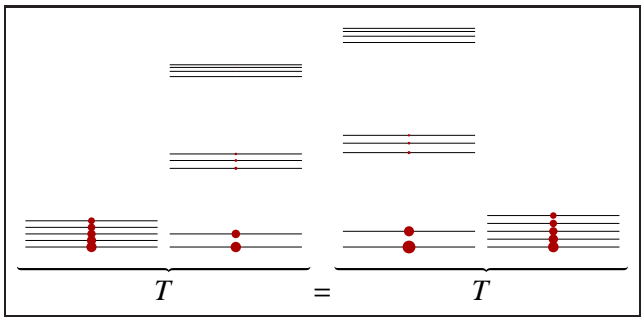}}
\caption{\label{fig:levelsconstT}Ananke's assignment of dot spectra, level occupation, and corresponding temperatures. Energy of each level is shown on an arbitrary scale, and the tensor-product decomposition of each dot into two subsystems is indicated by a separation of each dot's spectrum into two subspectra. Occupation of each level within~$\rho$ is indicated by the size of the corresponding red circle on that level in accord with Figure~\ref{fig:dothistogram}. For each dot, the subsystem with the evenly spaced spectrum is to become the local clock against which the dynamics of the other subsystem can be tested (see Sec.~\ref{subsec:clocks}). Ananke insists on Isothermality and thus requires that both dots be at the same temperature if they are both to be in thermal equilibrium with the laboratory. For Ananke, therefore, the two dots have spectra that are different (but still proportional to each other), with variations in energy levels explaining the differences in occupation of the corresponding levels in each dot. The clocks made from each dot run synchronously.}
\end{figure}

\begin{figure}[t!]
\centering
\fbox{\includegraphics[width=0.95 \columnwidth]{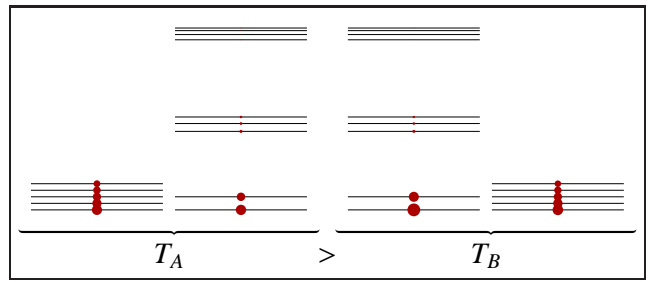}}
\caption{\label{fig:levelsdiffT}Chronos's assignment of dot spectra, level occupation, and corresponding temperatures. The tensor-product decomposition, occupations of each level, and subsystem assignment for each dot are the same as in Figure~\ref{fig:levelsconstT}, but the energies of each level are different from those in that Figure. Chronos chooses what he believes is a simpler explanation for the proportionality of the two dots' spectra: that the two dots are identical but at different temperatures. For Chronos, therefore, the two dots have the same spectrum, with variations in temperature explaining the differences in occupation of the corresponding levels in each dot. The clocks made from each dot run at rates in accord with the TE relation for the chosen temperatures (see Sec.~\ref{subsec:tempdiffTTH}).}
\end{figure}

Clearly the dots are similar, but their spectra are not exactly the same. Ananke and Chronos each explain the discrepancy in two different ways. Ananke supposes that since the dots were noted as being at ``room temperature'' they must each have been in a thermal state at the same temperature, resulting in the spectra of Figure~\ref{fig:levelsconstT}. Chronos disagrees based on the proportionality of their spectra, suggesting instead that a simpler explanation for the discrepancy is the two dots were \emph{identical} and just at slightly different temperatures. The spectra he assigns are shown in Figure~\ref{fig:levelsdiffT}. Since the dots themselves have been destroyed, they cannot tell who is right. They both agree, however, that the two subsystems comprising each individual dot must be at the same temperature because they were described in the notebook as being measurements made on a single dot. (Perhaps they were different internal degrees of freedom.) They then use the methods of Sec.~\ref{subsec:tempdiffTTH} to make predictions about the dynamics and obtain the TE relation as the way to relate them.

Without the dots at their disposal, they cannot test their predictions, so they email Chaos with their results and tell him they are at an impasse. He writes back that it's okay and that the real purpose of the dot experiment was to engineer an analog system for simulating equilibrium in a static gravitational field without using actual gravity. In fact, Ananke's assessment is correct (i.e.,~that the differences in the dots' spectra were due to a difference in their design). Furthermore, the dots themselves were not vandalized at all but were kept hidden so that the students would realize that \emph{Chronos's description is just as valid as Ananke's}. The lesson for the students is that, despite what is taught in thermodynamics textbooks, a priori there is no reason to insist on equal temperatures for composite systems in equilibrium as long as one uses the TTH to define relational clocks, as well. If one assigns different temperatures and uses the TTH to make relational predictions, then the empirical content of the TE relation appears automatically.

\subsection{Generality of this result}

Throughout this section, we have proceeded without reference to any ``objective'' time or Hamiltonian, and as a consequence we recovered a consistency relation resembling the TE relation for the use of the TTH locally.  We have, however, continued to make the assumption that it is known ``by independent means'' that the system to which we apply the TTH is in thermal equilibrium.  This is equivalent to ensuring ``by independent means'' that our clocks really are clocks, allowing us to maintain contact with known physics.

We note, finally, that this assumption was nowhere required to obtain the basic results of this section. In fact, the last vestiges of an independent time may finally be dropped, if so desired.  In fact, the original formulation of the TTH~\cite{Connes1994} is precisely this---i.e.,~that \emph{any} full-rank state~$\sigma$ is \emph{by definition} an equilibrium state with respect to the flow generated by $H_\sigma$.  Our results are thus a completely general feature of the TTH, with or without an independent means of verifying thermality.  It is in this fully general context that the TTH remains speculative, however. The extent to which thermal time represents a ``useful time'' in the absence of such additional assumptions (the original context for which the TTH was proposed) is simply not known.  Nevertheless, it is important to point out that our results continue to hold whether this final assumption of ``independently verified thermality'' is included or not.  If it is retained, our clock-construction protocol and consistency relation connects with known physics in the manner we have described above. If it is dropped, then our protocol and derived relation remain useful precisely to the extent that the TTH itself is discovered to be useful as a fundamental principle of physics.

\section{Discussion}
\label{sec:discussion}

Conceptually, the TTH is a simple idea: it says that one can define dynamics in terms of an equilibrium state instead of the reverse as is normally done. This has led, in the literature, to interesting results for expanding universes~\cite{Rovelli1993a}, the Unruh effect~\cite{Connes1994}, and the TE relation~\cite{Rovelli2011} in the context of a time flow emerging from the TTH. None of that work has used the principle of relationalism, however, to explore the empirical content of the predictions. In particular, while the work reported in Ref.~\cite{Rovelli2011} applies the TTH in the context of the TE relation, obtaining temperature as a ``speed of time,'' there was no operational meaning given to this term beyond that given by the notion of relativistic proper time itself. We have used the fact that temperature in the TTH is a free parameter to show that its role as a ``speed of time'' emerges, even in the absence of an actual gravitational field, when relational clocks are used. This effect is invisible at the local level but is visible (and functions like a redshift) when comparing relational clocks at different locations, as Chronos's choices in Sec.~\ref{subsec:tempdiffTTH} showed. In order to compare local clock readings, one can use a uniform temperature assignment to obtain a global thermal time coordinate, as Ananke did in that same section. The extension that this work provides beyond the results of Ref.~\cite{Rovelli2011} is that these effects occur in any context involving temperature variations within a global equilibrium state as long as relational clocks are used, not just in a static gravitational field. Using the TTH in this way predicts operational time dilation as a result of the temperature variation without invoking relativity explicitly. In fact, the analog experiment proposed in Sec.~\ref{subsec:exampletempvar} shows that actual gravitation is unnecessary for seeing this effect. Only global thermal equilibrium plus an assignment of different local temperatures is required.

The usual story is that gravity causes clocks to run more slowly (redshift) and that this causes the local temperatures to vary. We can write this schematically as
\begin{center}
time dilation $+$ therm eq\ $\to$ TE temperature variation,
\end{center}
where `TE' indicates that the temperatures vary commensurately with the time dilation in accord with the TE relation. In fact, this implication was shown to arise naturally from the TTH in Ref.~\cite{Rovelli2011} and from information erasure (Landauer's principle) in Ref.~\cite{Daffertshofer2007}. But the work presented here leads us to speculate about whether gravitational effects are fundamental in this relation between time and temperature. We wonder instead whether the notion of equilibrium contains this connection wholly within it. Specifically, what we have shown here is that we can reorder the terms in this implication, if we also include relationalism, so that
\begin{center}
therm eq\ $+$ temp var $+$ relationalism $\to$ TE time dilation.
\end{center}
Temperature becomes a free parameter, whose variation among an extended sample in equilibrium induces a compensating dilation in time---effectively, a redshift---in accord with the TE relation.

This is reminiscent of Feynman's hot plate example~\cite{Feynman1963v2}, which is based originally on a similar example by Poincar\'e~\cite{Poincare1905}. In Feynman's example, an ant walking on a hot plate with a radial temperature gradient experiences curved spatial geometry because of its interpretation of thermal expansion effects as length contraction. Our work is similar but for time instead of space. The crucial ingredient to both Feynman's example and our own work is \emph{relationalism}. The ant uses physical rods to measure length, not mere coordinates, and we require the construction of physical clocks whose dynamics are also predicted by the same method used to predict those of the system(s) in question---i.e.,~by using the TTH.

The example of the quantum dots in Sec.~\ref{subsec:exampletempvar} is not as far fetched as it may seem. One might be tempted to think that since the dots ``really are'' created to be different and that the temperatures ``really are'' the same that there ``really is'' no time dilation. At this point it helps to recall Sec.~\ref{subsec:objectivity}. What does it mean to assert something is really the case over another interpretation? Specifically, what would happen if all an observer had were relational clocks, rulers, and other objects made from items appearing to have spectra in one region of her world that are just like the spectra of similar items in another region except for an overall scaling factor? Wouldn't observers in that world be justified in assuming that these items really are the same and that differences in occupation numbers of the equilibrium state are a result of varying temperatures instead of different ``stuff?''

To see more clearly that the answer is `yes,' consider that in modern astronomy an absorption spectrum is often observed that would be from hydrogen except that all the energies have been scaled by a given factor. Is a new element called ``squished hydrogen'' proposed to explain this? No. A much simpler explanation is that hydrogen atoms here are the same as hydrogen atoms there and that the discrepancy is explained by a redshift. (This simplicity remains, under certain assumptions, even if we were to have no knowledge of relativity beforehand.) Operationally, this is exactly the same thing that Chronos did with his observation of the two dots' spectra. \emph{Our belief that the laws of physics---and the constituents of the universe---are the same everywhere could lead us to postulate redshifts to explain the scaling of observed atomic spectra, even if relativity had not been discovered yet.}

What we have shown is that an effective redshift (time dilation) occurs regardless of whether the scaling of the spectra is really caused by gravity or not. We have said nothing about why the spectra in one region look proportional to those in another region, and this remains an open question. What we have shown, however, is that no matter what temperature variation an agent assigns to a sample in equilibrium, if she also constructs local relational clocks using dynamics obtained by the TTH, then she will observe time dilation in accord with the TE relation for the temperatures that have been assigned, regardless of what those temperatures are or what explanation is given for the variation.

\acknowledgments

We thank Robin Blume-Kohout, Sarah Croke, Matteo Smerlak, Eric Cavalcanti, Daniel Gottesman, Andrzej Dragan, and Tim Ralph for discussions. N.C.M.\ and S.J.O.\ thank The University of Queensland and the Foundational Questions Institute (FQXi) for financial and institutional support. N.C.M.\ and G.J.M.\ thank the organizers and attendees of the 2011 Relativistic Quantum Information Workshop at The University of KwaZulu-Natal for a forum to present an earlier version of this work and for the useful feedback that resulted. Similar thanks goes to the attendees of the Relativistic Quantum Information Workshops, organized by Tim ralph, in 2009 and 2010 at The University of Queensland. Research at Perimeter Institute is supported by the Government of Canada through Industry Canada and by the Province of Ontario through the Ministry of Research \& Innovation.


\bibliographystyle{bibstyleNCM}
\bibliography{TolmanTTH}

\end{document}